\documentstyle[12pt,aaspp4]{article}

\def\t0{\theta_{\circ}}

\def\be{\begin{equation}}
\def\en{\end{equation}}

\begin{document}

%\shorttitle{Young Brown Dwarf Disk Census}
%\shortauthors{Jayawardhana et al.}

\title{A Disk Census for Young Brown Dwarfs}

\author{Ray Jayawardhana}
\affil{Department of Astronomy, University of Michigan, 830 Dennison Building, Ann Arbor, MI 48109, U.S.A.}
%\email{rayjay@umich.edu}

\author{David R. Ardila}
\affil{Bloomberg Center for Physics and Astronomy, Johns Hopkins University, 3400 N. Charles Street, Baltimore, MD 21218, U.S.A.}

\author{Beate Stelzer}
\affil{Osservatorio Astronomico di Palermo, Piazza del Parlamento 1, I-90134 Palermo, Italy}

\and

\author{Karl E. Haisch, Jr.}
\affil{Department of Astronomy, University of Michigan, 830 Dennison Building, Ann Arbor, MI 48109, U.S.A.}

\begin{abstract}
Recent surveys have identified sub-stellar objects down to 
planetary masses in nearby star-forming regions. Reliable 
determination of the disk frequency in young brown dwarfs is of 
paramount importance to understanding their origin. Here we report 
the results of a systematic study of infrared $L^\prime$-band (3.8$\mu$m) 
disk excess in $\sim$50 spectroscopically confirmed objects near 
and below the sub-stellar boundary in several young clusters. 
Our observations, using the ESO Very Large Telescope, Keck I  
and the NASA Infrared Telescope Facility, reveal that a significant 
fraction of brown dwarfs harbor disks at a very young age. Their 
inner disk lifetimes do not appear to be vastly different from those of 
disks around T Tauri stars. Our findings are consistent with the 
hypothesis that sub-stellar objects form via a mechanism similar 
to solar-mass stars. 
\end{abstract}


\section{Introduction}
The current paradigm of low-mass star formation holds that a young star 
accretes material from a circumstellar disk during the first (few) million  
years of its lifetime. The disk also provides the building material for
planets. Over the past two decades, substantial observational evidence 
has accumulated to support this picture. However, much of that evidence 
rests on studies of stars within a relatively narrow mass range. In 
particular, there are few observational constraints on the formation of 
objects near or below the sub-stellar boundary. 

The past five years have seen the detection of a large number of 
sub-stellar objects in the solar neighborhood and in young clusters. 
Large-scale optical and near-infrared surveys of the sky --such as the 
Sloan Digital Sky Survey and the 2-Micron All-Sky Survey (2MASS)-- have turned 
up (older) field brown dwarfs with L and T spectral types (e.g., 
Kirkpatrick et al. 2000; Fan et al. 2000). Meanwhile, their young 
counterparts have been identified in star-forming regions through more 
targeted searches (e.g., Luhman et al. 2000; Ardila, Mart\'\i n \& Basri 
2000). Perhaps the most surprising of recent findings has been the 
discovery of a population of young, isolated planetary mass objects 
--or ``planemos''-- in the $\sigma$ Orionis 
(Zapatero-Osorio et al. 2000), Trapezium (Lucas \& Roche 2000) and
Upper Scorpius (Mohanty, Jayawardhana \& Basri 2003) regions. 

Still, the formation mechanism for brown dwarfs is open to debate. 
Reipurth \& Clarke (2001) and Bate et al. (2002) have proposed
that brown dwarfs are stellar embryos, ejected from newborn multiple 
systems before they can accrete sufficient mass for hydrogen fusion. 
Padoan \& Nordlund (2003), on the other hand, have suggested that 
brown dwarfs form in the same way as more massive stars, via `turbulent 
fragmentation'.

Studies of disks around young brown dwarfs provide one of the few 
observational probes to distinguish among these scenarios. If a 
substantial fraction of brown dwarfs harbor large, long-lived 
accretion disks, the implication is that extremely low-mass objects 
may form via a mechanism similar to higher mass stars. Thus, reliable 
determination of the disk frequency and lifetime in young brown dwarfs 
is of paramount importance to our 
understanding of the origin and diversity of sub-stellar objects down 
to planetary masses. In a pioneering study, Muench et al. (2001) reported
that 65\%$\pm$15\% of brown dwarf {\it candidates} in the Trapezium 
cluster exhibit JHK colors consistent with circumstellar disks. 

Here we report on a systematic study of infrared $L^\prime$-band (3.8$\mu$m) 
disk excess in a large sample of {\it spectroscopically confirmed} objects 
near and below the sub-stellar boundary in several nearby star-forming 
regions. Our longer-wavelength observations are much better at detecting
disk excess above the photospheric emission and are less susceptible to the 
effects of disk geometry and extinction corrections than $JHK$ studies
(e.g., Haisch, Lada \& Lada 2001; Liu, Najita \& Tokunaga 2003). 

\section{Observations}
Our target sample consists of very low mass (VLM) objects with known 
spectral types later than M5 in nearby star-forming regions. They are drawn 
from surveys of
$\rho$ Ophiuchus (Wilking, Greene \& Meyer 1999), IC 348 (Luhman 1999), 
Chamaeleon I (Comer\'on, Neuh\"auser \& Kaas 2000; Comer\'on, Rieke \& 
Neuh\"auser 1999), Taurus (Brice\~no et al. 2002), Upper Scorpius 
(Ardila, Mart\'\i n \& Basri 2000), $\sigma$ Orionis (B\'ejar, Zapatero 
Osorio \& Rebolo 1999), and the TW Hydrae Association (Gizis 2002). 

Observations were carried out at three different telescopes during 2002. 
Table 1 provides a summary log. When $JHK$ magnitudes were not available 
in the literature or in the 2MASS database, we also obtained photometric 
observations in those filters. The transformations between the different 
photometric systems are smaller than the measuring errors (Carpenter 2001). 

The observations at the Very Large Telescope (VLT) were obtained in 
service mode using the Infrared Spectrograph and Array Camera (ISAAC; 
Moorwood et al. 1998).  At Keck I, we used the Near InfraRed Camera (NIRC; 
Matthews \& Soifer 1994). $JHK$ data were obtained in a five-point pattern. 
The $L^\prime$-band data were acquired in the chop-nod mode (15 to 
30$\arcsec$ offsets), in order to cancel out the variable thermal background.  
Each observation consisted of a large number of co-added frames with 
short exposure times. Observations of several $\rho$ Oph sources were 
obtained with the 
NSFCAM camera (Rayner et al. 1993; Shure et al. 1994) on the 
NASA InfraRed Telescope Facility (IRTF). Each source was observed in a five 
point dither pattern. At each dither position, the telescope was nodded to 
separate sky positions 30\arcsec~ of the target observation.

All the data were reduced using standard tools within the Image Reduction 
and Analysis Facility 
(IRAF)\footnote[1]{IRAF is distributed by the National Optical Astronomy 
Observatories, which are operated by the Association of Universities for 
Research in Astronomy, Inc., under cooperative agreement with the National 
Science Foundation.}. Flat fields were obtained by combining individual 
images. 
For the $JHK$ observations the sky was subtracted from each frame. 
Aperture photometry was performed for each source, using the PHOT routine 
within IRAF. The standards were observed on the same nights 
and through the same range of airmasses as the target sources. Zero 
points and extinction coefficients were established for each night. 
The photometric accuracy of our observations are typically $\pm$ 
0.10 magnitudes, though in a few cases they can be $\pm$ 0.20 magnitudes.

\section{Results and Discussion}
Table 2 lists the $JHKL^\prime$ magnitudes and $J-H, H-K, K-L^\prime$ 
colors of our targets. Figure 1 shows their $K-L^\prime$ colors as 
a function of spectral type in comparison to the locus of field M dwarfs 
from Leggett et al. (2002). It is clear that a significant fraction of our 
targets have redder $K-L^\prime$ colors than would be expected from 
photospheric emission alone. The lower envelope of their color distribution  
does agree reasonably well with the field M dwarf locus. 

In Figure 2, we plot the $J-H$/$K-L^\prime$ color-color diagram for
our target sample. For comparison, we also plot the empirical loci of 
colors for giants and for main-sequence dwarfs from Bessell \& Brett 
(1988) and Leggett et al. (2002). Again, we find that a large fraction of
our sources fall to the right of the reddening band for M dwarfs and 
into the infrared excess region of the color-color diagram. 

Since all objects in our sample have known spectral types, we can 
determine their intrinsic photospheric $K-L^\prime$ by comparison to 
field M dwarfs of the same type and by using a reddening law (e.g., 
Cohen et al. 1981). An excess of $K-L^\prime \gtrsim$ 
0.2 above the stellar photosphere is usually indicative of an optically 
thick disk around a late-type object and is also a reasonable criterion 
given the typical photometric errors in our data as well as possible 
color differences in field objects due to higher gravity/older age. 
Table 3 lists the spectral 
types and $K-L^\prime$ excesses of our targets, along with the equivalent
width of their H$\alpha$ emission, when available in the literature. 
Table 4 gives the infrared excess fraction in each star-forming region.  

Of the seven $\rho$ Ophiuchus objects for which we have $L^\prime$ 
photometry, four exhibit $K-L^\prime$ excesses. Two other objects on 
our target list --GY 11 and GY 141-- which we could not observe at 
$L^\prime$ also show evidence of optically thick disks according to 
the Infrared Space Observatory (ISO) measurements at 6.7$\mu$m and 
14.3$\mu$m (Testi et al. 2002; Comer\'on et al. 1998). The high disk 
fraction among $\rho$ Oph brown dwarfs is not surprising, given
that $\rho$ Oph is the most embedded, and probably the youngest, cluster 
in our survey at an age of $\lesssim$1 Myr. Our finding is also consistent 
with that of Natta et al. (2002) who reported ISO-detected mid-infrared 
excess in nine other VLM sources in this region. These authors were able 
to fit the excesses with irradiated disk models. 

We find disk fractions of 40\%--60\% in IC 348, Chamaeleon I, Taurus and 
Upper Scorpius regions. Based on ISO observations, Natta \& Testi (2001) 
have already shown that Cha H$\alpha$ 1, ChaH$\alpha$ 2 and ChaH$\alpha$ 9 
harbor mid-infrared spectral energy distributions consistent with the presence
of dusty disks. ChaH$\alpha$ 2, which shows a large $K-L^\prime$ excess 
(0.97 mag) in our data is a probable close ($\sim$0.2'') 
binary with roughly equal-mass companions (Neuh\"auser et al. 2002). 
It is possible that a few of our targets harbor infrared companions that
contribute to the measured excess, but this is unlikely in most cases. 
The disk fractions we report for IC 348 and Taurus are lower than those 
found by Liu, Najita \& Tokunaga  (2003). This is primarily because we use 
a more conservative criterion of $K-L^\prime >$ 0.2 for the presence of 
optically thick disks whereas Liu et al. consider all objects with 
$K-L^\prime >$ 0 as harboring disks. In IC 348, our disk fraction is 
comparable to that derived from H$\alpha$ accretion signatures in 
high-resolution optical spectra (Jayawardhana, Mohanty \& Basri 2003; 
White \& Basri 2002). However, in Taurus and Upper Sco, which may be 
slightly older at $\sim$3-5 Myrs, we find $K-L^\prime$ excess 
in $\sim$50\% of the targets whereas only three of out 14 Taurus VLM 
objects and one out of 11 Upper Sco sources exhibit accretion-like H$\alpha$ 
(Jayawardhana, Mohanty \& Basri 2002; 2003). This latter result suggests 
that dust disks may persist after accretion has ceased or been reduced to 
a trickle, as also suggested by Haisch, Lada \& Lada (2001). 

In the somewhat older ($\sim$5 Myr) $\sigma$ Orionis cluster, only about 
a third of the targets show infrared excess. Neither of the two brown 
dwarf candidate members of the $\sim$10-Myr-old TW Hydrae association 
(Gizis 2002) shows excess. Gizis (2002) reported strong H$\alpha$ emission 
(equivalent width $\approx$ 300 \AA) from one of the TW Hydrae objects, 
the M8 dwarf 2MASSW J1207334-393254, and suggested it could be due to 
either accretion or chromospheric activity. Given the 
lack of measurable $K-L^\prime$ excess in this object, accretion 
now appears less likely as the cause of its strong H$\alpha$ emission. 
Our findings in $\sigma$ Ori and TW Hya associations, albeit for a small 
sample of objects, could mean that the inner disks are clearing out 
by the age of these groups. Similar results have been found for 
T Tauri stars in the TW Hydrae association (Jayawardhana et al. 1999).  

Jayawardhana, Mohanty \& Basri (2003) report a decrease in the fraction 
of accreting brown dwarfs with increasing age in a study of H$\alpha$ line
profiles in high-resolution optical spectra. This trend is in general 
agreement with our results here based on infrared excess measurements. 
However, as shown in Figure 3, we do not find a one-to-one correlation 
between objects with $K-L^\prime$ excess and those with large H$\alpha$ 
equivalent widths. As mentioned earlier, several of the Taurus and 
Upper Sco VLM objects
with excess do not exhibit accretion-like H$\alpha$. While this is somewhat
surprising, it is not unprecedented. For example, in IC 348, Haisch, Lada,
\& Lada (2001) found that 10/17 weak-line (H$\alpha$ EW $<$ 10\AA) T Tauri 
stars in their sample showed $K-L^\prime$ excess. Persistence of disks after 
the accretion rates have significantly diminished could explain these
results. Another complication is that the H$\alpha$ equivalent widths in 
Table 3 are drawn from a variety of sources in the literature. Many of the
measurements have been made with low- or medium-resolution spectra. 
It has been shown that low-resolution spectra of late-type objects 
systematically overestimate the H$\alpha$ equivalent width as a result of 
blending with the 6569\AA~ TiO band-head (e.g., Tinney \& Reid 1998). 
This effect can confuse the relation between H$\alpha$ width and infrared 
excess. 

\section{Concluding Remarks}
Our results, and those of Muench et al. (2001), Natta et al. (2002), and 
Liu, Najita \& Tokunaga (2003) show that a large fraction of very young brown 
dwarfs harbor near- and mid-infrared excesses consistent with dusty disks. 
High-resolution optical spectra have revealed accretion signatures 
in many sources (Jayawardhana, Mohanty \& Basri 2003). Taken together, the 
evidence is compelling that sub-stellar objects are surrounded by accretion 
disks similar to those around T Tauri stars, possibly less massive. While the 
samples are still relatively small, the timescales for inner disk depletion 
do not appear to be vastly different between brown dwarfs and T Tauri stars
(whereas Armitage \& Clarke 1997, for example, predict faster disk 
evolution for ejected objects). Therefore, the evidence to date is 
consistent with a common formation scenario for low-mass stars and brown 
dwarfs. Far-infrared observations with
the {\it Space InfraRed Telescope Facility} and/or the {\it Stratospheric 
Observatory For Infrared Astronomy} will be crucial for deriving the
sizes of circum-sub-stellar disks and providing a more definitive test of the
ejection hypothesis for the origin of brown dwarfs.

\acknowledgements
We thank Geoff Marcy for useful discussions, constant encouragement and
access to Keck and Kevin Luhman for valuable assistance during the December 
2002 Keck run. We are grateful to the staff members of the VLT, 
Keck and IRTF observatories for their outstanding support. We also thank 
Fernando Comer\'on and staff at the ESO User Support Group for their prompt 
responses to our queries. We would like to acknowledge the great cultural 
significance of Mauna Kea for native Hawaiians, and express our gratitude 
for permission to observe from its summit. This work was supported in part 
by NSF grant AST-0205130 and NASA grant NAG5-13136 to R.J.

%\newpage

\clearpage
\begin{figure}
%\epsscale{1.0}
\plotone{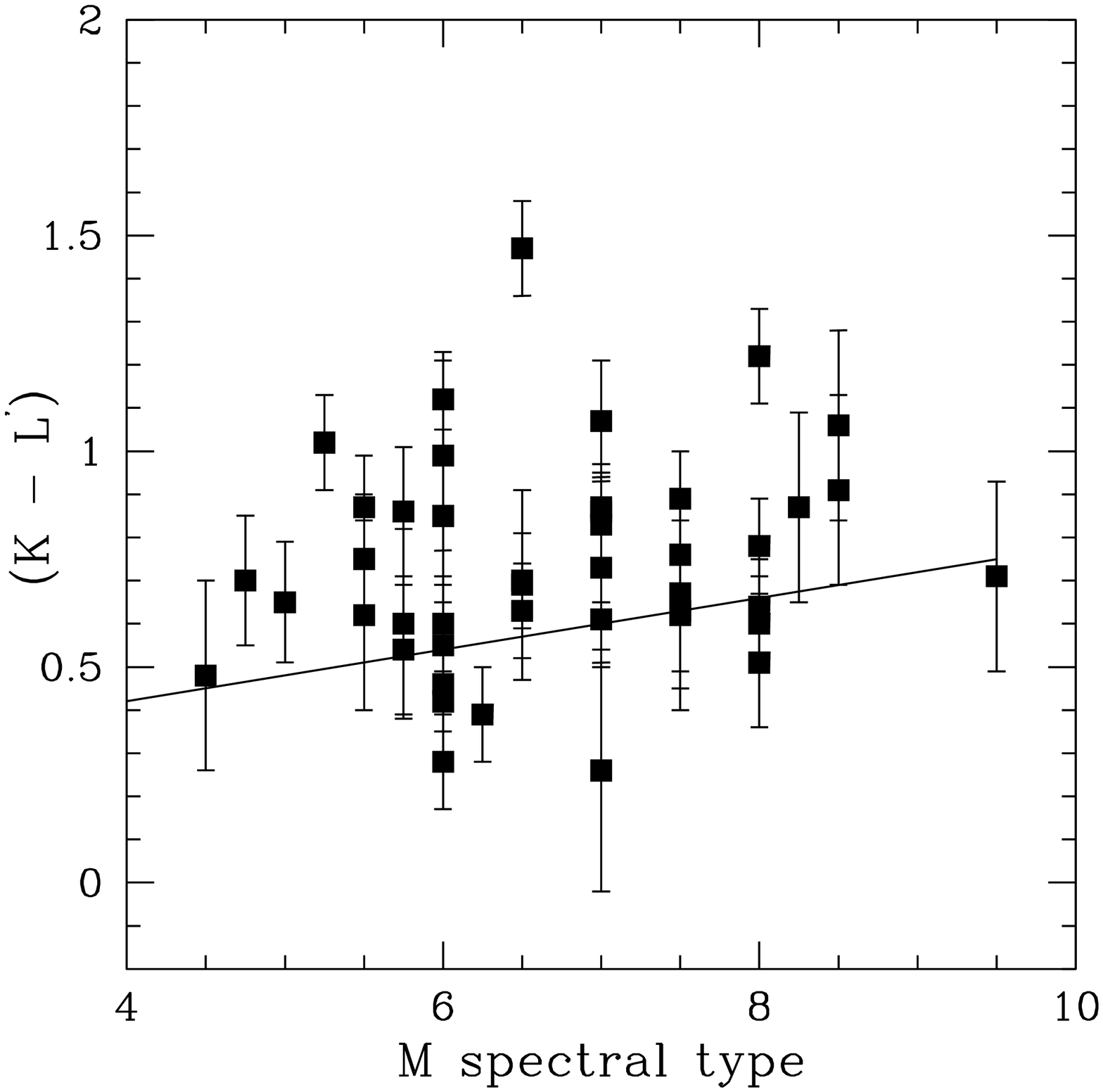}
\caption{$K-L^\prime$ color as a function of spectral type.}
\end{figure}

\clearpage
\begin{figure}
%\epsscale{0.5}
\plotone{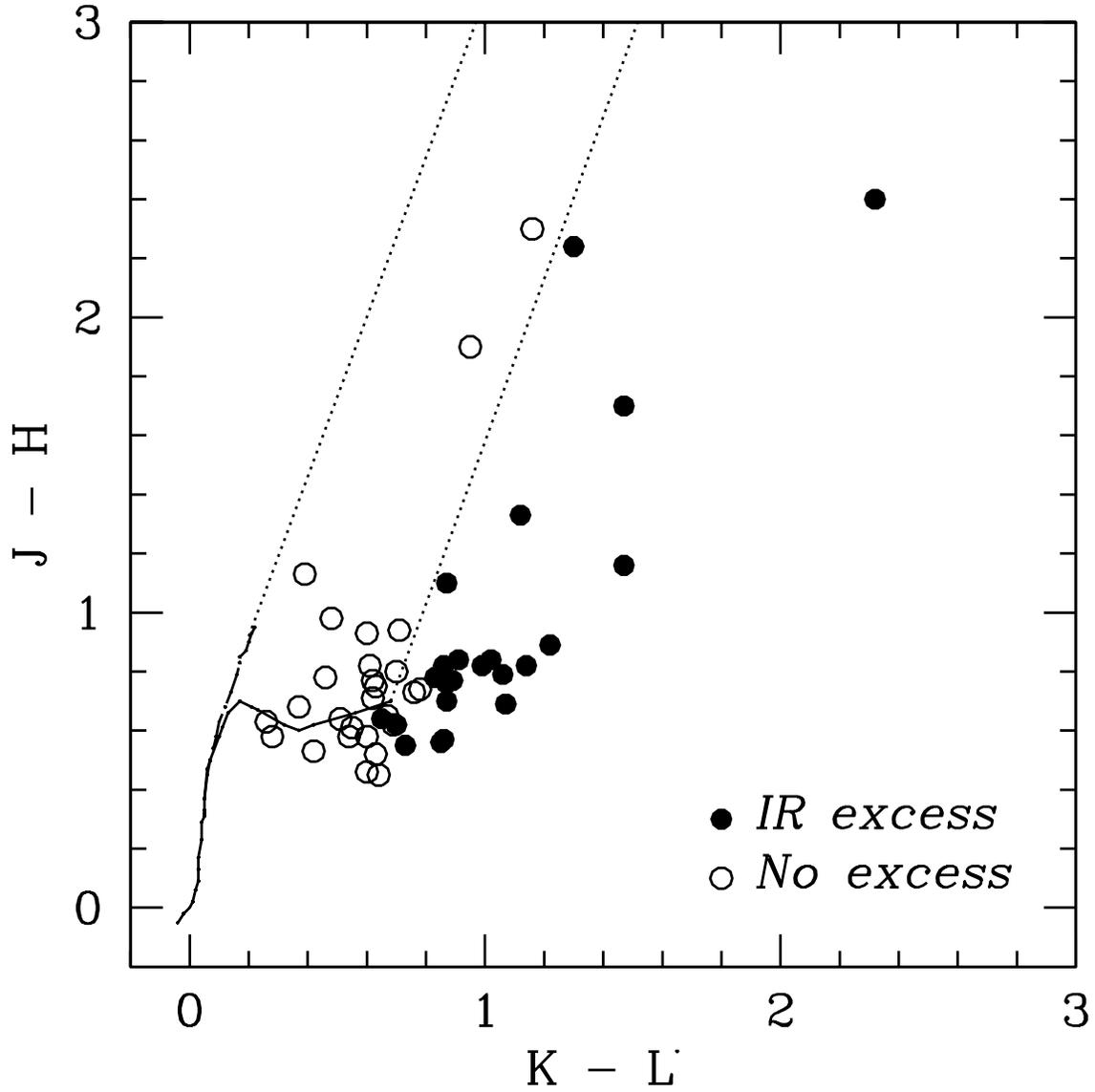}
\caption{$J-H$/$K-L^\prime$ color-color diagram for our target sample. Also 
plotted are the empirical loci of colors for giants (solid) and for main-sequence 
dwarfs (dashed) from Bessell \& Brett (1988) and Leggett et al. (2002) and the 
reddening vectors (dotted). The filled circles are stars with  $E(K-L^\prime) >$ 
0.2.}
\end{figure}

\clearpage
\begin{figure}
%\epsscale{0.5}
\plotone{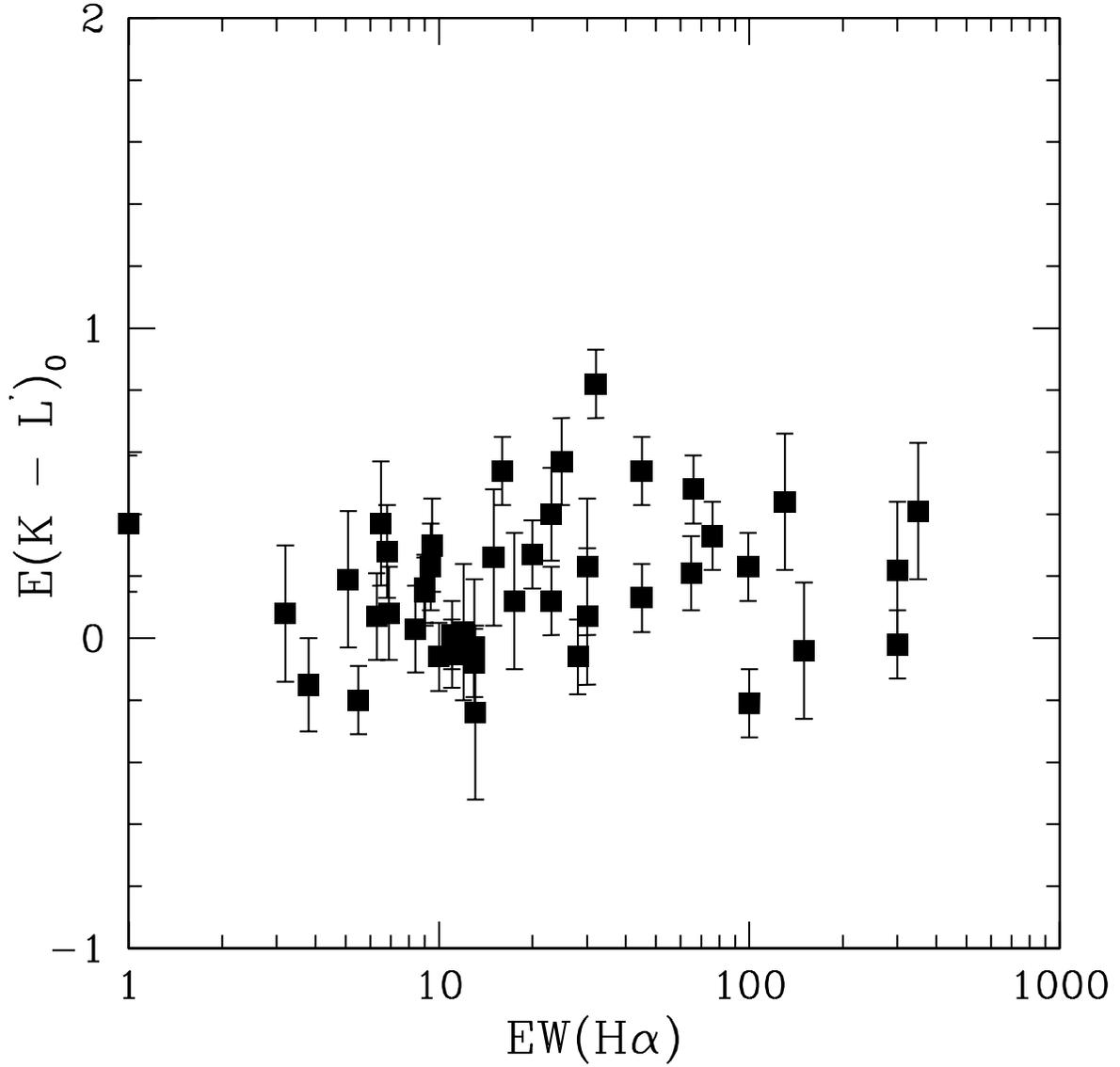}
\caption{Relation between $K-L^\prime$ excess and equivalent width of 
H$\alpha$ emission in Angstroms, from the literature, as listed in Table 3.} 
\end{figure}

\clearpage
%\documentstyle[12pt,preprint]{aastex}
%\begin{document}
\begin{deluxetable}{lcccccc}
\footnotesize
\tablecaption{Log of Observations
\label{table1}}
\tablewidth{0pt}
\tablehead{UT Date & Telescope & Target & Filter & On-Source Integration (s) & Flux Standard }
\startdata

2002 January 11 &   VLT-ANTU  &  S Ori 12 &     L'   &     240     &           HR4638\nl
                &             &  S Ori 17 &     L'   &     720     &            HR4638\nl
                &             &  S Ori 25 &     L'   &     1080    &          HR4638\nl
                &             &   S Ori 29&     L'   &     780   &         HR4638\nl
                &             &  S Ori 39 &    L'    &     1860   &           HR4638\nl
                &             &   S Ori 40 &     L'  &      1320  &             HR4638\nl
2002 April 15   &   VLT-ANTU  &    Cha H$\alpha$ 1   &   L' &        118.8      &                 HD83844\nl
                        &      &   Cha H$\alpha$ 2   &          L'     &          118.8         &              HD83844\nl
                        &       &   Cha H$\alpha$ 3  &           L'      &         118.8        &               HD83844\nl
                        &       &   Cha H$\alpha$ 4   &          L'       &        118.8       &                HD83844\nl
                        &       &   Cha H$\alpha$ 5   &          L'      &         118.8       &                HD83844\nl
                        &       &   Cha H$\alpha$ 6   &          L'      &         118.8       &                HD83844\nl
                        &        &   Cha H$\alpha$ 7   &          L'       &        118.8        &               HD83844\nl
                        &       &   Cha H$\alpha$ 8   &          L'       &        118.8        &               HD83844\nl
                        &       &   Cha H$\alpha$ 9    &         L'       &        118.8        &               HD83844\nl
                        &      &    Cha H$\alpha$ 10   &          L'     &         297          &            HD83844\nl
                        &       &    Cha H$\alpha$ 11   &          L'     &         594           &             HD83844\nl
                        &      &   Cha H$\alpha$ 12   &          L'     &         118.8         &             HD83844\nl
2002 April 20   &     VLT-ANTU    &  UScoCTIO 104  &      L'      &        118.8      &                HD130163\nl 
%                          &    &    UScoCTIO 104  &      L'      &        118.8      &                HD130163\nl
                         &    &    UScoCTIO 109  &      L'      &        237.6      &               HD130163\nl
%                         &   &   UScoCTIO 121  &      L'  &      415.8    &                 HD130163\nl
                        &    &    UScoCTIO 128  &      L'      &        415.8      &               HD130163\nl
2002 April 23   &     Keck I   &      GY 202     &       K     &          6       &                 HD129655\nl
                &              &                 &       L'    &          10      &                 HD129655\nl
2002 April 24   &     Keck I    &      UScoCTIO 112  &K &        35   &              LHS2397a\nl
                &              &                  & L' &       60          &       LHS2397a        \nl                                             
                &              &    UScoCTIO 130 & K      &   35           &      LHS2397a\nl
                &              &                &L'    &    300          &         LHS2397a \nl  
                &              &     UScoCTIO 137 & K  &       10           &         LHS2397a\nl
                &              &               &  L' &       300          &          LHS2397a \nl                                  
                &              &    2MASSW J1207334-393254  &  J  &       25           &         LHS2397a\nl
                &              &               &  H &        25           &         LHS2397a\nl
                &              &               &  K  &       35           &         LHS2397a\nl
                &              &               &  L' &       40           &         LHS2397a\nl
                &              &    2MASSW J1139511-315921   &  J  &       50           &         LHS2397a\nl
                &              &             & H     &    50            &        LHS2397a\nl
                &              &             & K     &    35            &        LHS2397a\nl
                &              &            & L'     &   40            &        LHS2397a\nl
2002 May 30     &     VLT-ANTU  &   UScoCTIO 100 & J &        80                   &  S860-D\nl
                           &    &              & H &        80                   &  S860-D\nl
                          &     &             &  K &        80                  &   S860-D\nl
                         &     &   UScoCTIO 104 & J    &     80               &      S860-D\nl
                          &      &             & H    &     80                 & S860-D\nl
                          &      &             & K    &     80              &       S860-D  \nl                            
                           &     &   UScoCTIO 109 & J &        80                   &  S860-D\nl
                           &     &             & H &        80                   &  S860-D\nl
                           &     &             & K &        80                   &  S860-D\nl
%                          &     &   UScoCTIO 121 & J    &     80              &       S860-D\nl
%                          &      &             & H    &     80              &       S860-D\nl
%                          &      &             & K    &     80                 &    S860-D \nl                   
                         &     &         UScoCTIO 128      & K &        80                &     S860-D\nl
                         &     &    UScoCTIO 132 & J &        80                &     S860-D\nl
                          &     &              &  H &        80               &      S860-D\nl
                          &    &               & K &        80                   &     S860-D\nl
2002 June 12 &      VLT-ANTU    &     CHXR 78c &   L' &       118.8            &      S587-T\nl
                            &  &   UScoCTIO 100 & L' &      118.8                &  S587-T\nl
2002 June 13  &      VLT-ANTU  &      CHXR 73 &     L' &        118.8              &       S860-D\nl
                           &    &   CHXR 74     & L' &       118.8               &   HR4638\nl
                           &    &   Cha H$\alpha$ 12    & J &        80                   &  S860-D\nl
                          &      &             & H &        80                   &  S860-D\nl
                         &      &        & K &        80                   &  S860-D\nl
2002 June 14   &     VLT-ANTU   &     Cha H$\alpha$ 8   &   J &        80           &          S860-D\nl
                            &     &            & H &        80                &     S860-D\nl
                         &      &        & K &        80                   &  S860-D\nl
2002 June 15   &      IRTF      &     CRBR 14   &  J &          25  & HD161903\nl
                        &         &          & H &         25 & HD161903\nl
                         &         &           & K &         25 & HD161903\nl
                         &         &           & L' &         25 & HD161903\nl
                        &       &   CRBR 15    & J &         25 & HD161903\nl
                        &        &            &  H &         25 & HD161903\nl
                        &       &             &  K &         25 & HD161903\nl
                        &       &             &  L' &         25 & HD161903\nl
                 &           &    GY 5      &  J &         25 & HD161903\nl
                       &       &              &  H &         25 & HD161903\nl
                        &      &              &  K &         25 & HD161903\nl
                      &         &             &  L' &         25 & HD161903\nl
                 &           &    GY 10     &  J &         25 & HD161903\nl
                      &         &             &  H &         25 & HD161903\nl
                      &         &             &  K &         25 & HD161903\nl
                      &        &              &  L' &         25 & HD161903\nl
                 &           &    GY 59     &  J &         25 & HD161903\nl
                      &         &             &  H &         25 & HD161903\nl
                      &         &             &  K &         25 & HD161903\nl
                       &        &             &  L' &         25 & HD161903\nl
                 &            &   GY 64     &  J &         25 & HD161903\nl
                       &         &             &  H &         25 & HD161903\nl
                       &         &             & K &         25 & HD161903\nl
                        &        &            &  L' &         25 & HD161903\nl                             
2002 August 12   &   VLT-ANTU  &      UScoCTIO 132&  L' &       475.2          &        HR6070 \nl
2002 December 19  &  Keck I   &    IC348 165     &  K &        5             &         GD50 \nl
                  &           &                  &  L' &       20                    & HD22686 \nl
                     &        &   IC348 256     & K &        5                &      GD50 \nl
                     &        &                & L' &       20                   &  HD22686 \nl
                     &       &    IC348 286     & K &        5                &      GD50 \nl
                     &       &                 &  L' &       100                 &    HD22686 \nl
                     &        &   IC348 363     & K &         5               &       GD50 \nl
                     &        &             &   L' &       400             &        HD22686 \nl
                     &       &    IC348 367    &  K &         5                 &     GD50 \nl
                    &        &                 &  L' &       400                &     HD22686 \nl
                     &       &    IC348 478     & K &         5                     & GD50 \nl
                   &         &                  & L' &       400                    & HD22686 \nl
2002 December 20  &  Keck I    &     KPNO-Tau 1    &      L' &         200         &            HD22686 \nl
           &           &           KPNO-Tau 2      &   L' &        80              &       HD22686 \nl
                  &           &        KPNO-Tau 3         &   L' &        20              &       HD22686 \nl
                   &       &        KPNO-Tau 4        &   L' &       200                 &    HD22686 \nl
                         &         &         KPNO-Tau 5       &   L' &        20               &      HD22686 \nl
                   &       &              KPNO-Tau 6  &   L' &       200                &     HD22686 \nl
                  &         &         KPNO-Tau 7       &   L' &       200                  &   HD22686 \nl
                  &         &          KPNO-Tau 8       &  L' &        20                  &   HD22686     \nl       
                   &           &            KPNO-Tau 9      &  L' &       600                  &   HD22686  
\enddata
\end{deluxetable}
%\end{document}

\clearpage
\begin{deluxetable}{lccrrrrrrr}
\footnotesize
\tablecaption{Positions and $JHKL^\prime$ Magnitudes and Colors
\label{table9}}
\tablewidth{0pt}
\tablehead{Source & RA(J2000) & Dec(J2000) & J\tablenotemark{a} & H\tablenotemark{a} & K\tablenotemark{a} & L$^\prime$ & (J-H) & (H-K) &
(K-L$^\prime$)}
\startdata
IC348 165 & 03 44 35.43 & $+$32 08 54.4 & 13.15 & 12.31 & 11.84 & 10.82 & 0.84 & 0.47 & 1.02\nl
IC348 256 & 03 43 55.14 & $+$32 07 55.0 & 13.59 & 13.02 & 12.54 & 11.68 & 0.57 & 0.48 & 0.86\nl
IC348 286 & 03 45 06.80 & $+$32 09 26.8 & 13.76 & 13.18 & 12.71 & 12.17 & 0.58 & 0.47 & 0.54\nl
IC348 363 & 03 44 17.00 & $+$32 00 15.3 & 14.83 & 14.19 & 13.80 & 13.29 & 0.64 & 0.39 & 0.51\nl
IC348 367 & 03 43 59.03 & $+$32 05 57.9 & 14.72 & 14.10 & 13.59 & 12.89 & 0.62 & 0.51 & 0.70\nl
IC348 478 & 03 44 35.97 & $+$32 11 15.9 & 16.17 & 15.04 & 14.60 & 14.21 & 1.13 & 0.44 & 0.39\nl
KPNO-Tau 1 & 04 15 14.72 & $+$28 00 09.5 & 15.09 & 14.25 & 13.74 & 12.83 & 0.84 & 0.51 & 0.91\nl
KPNO-Tau 2 & 04 18 51.15 & $+$28 14 33.3 & 13.89 & 13.18 & 12.74 & 12.12 & 0.71 & 0.44 & 0.62\nl
KPNO-Tau 3 & 04 26 29.38 & $+$26 24 14.2 & 13.32 & 12.50 & 12.08 & 11.09 & 0.82 & 0.42 & 0.99\nl
KPNO-Tau 4 & 04 27 28.01 & $+$26 12 05.3 & 14.98 & 14.04 & 13.31 & 12.60 & 0.94 & 0.73 & 0.71\nl
KPNO-Tau 5 & 04 29 45.68 & $+$26 30 46.7 & 12.61 & 11.96 & 11.50 & 10.83 & 0.65 & 0.46 & 0.67\nl
KPNO-Tau 6 & 04 30 07.25 & $+$26 08 20.7 & 15.00 & 14.21 & 13.66 & 12.60 & 0.79 & 0.55 & 1.06\nl
KPNO-Tau 7 & 04 30 57.21 & $+$25 56 40.0 & 14.50 & 13.80 & 13.22 & 12.35 & 0.70 & 0.58 & 0.87\nl
KPNO-Tau 8 & 04 35 41.85 & $+$22 34 11.6 & 12.96 & 12.38 & 12.00 & 11.40 & 0.58 & 0.38 & 0.60\nl
KPNO-Tau 9 & 04 35 51.43 & $+$22 49 11.9 & 15.49 & 14.67 & 14.17 & 13.03 & 0.82 & 0.50 & 1.14\nl
S Ori 12 & 05 37 57.40 & -02 38 45.0 & 14.20 & 13.64 & 13.28 & 12.43 & 0.56 & 0.36 & 0.85\nl
S Ori 17 & 05 39 04.40 & -02 38 35.0 & 14.77 & 14.19 & 13.79 & 13.51 & 0.58 & 0.40 & 0.28\nl
S Ori 25 & 05 39 08.80 & -02 39 58.0 & 14.67 & 14.15 & 13.76 & 13.13 & 0.52 & 0.39 & 0.63\nl
S Ori 29 & 05 38 29.50 & -02 25 17.0 & 14.83 & 14.30 & 13.96 & 13.54 & 0.53 & 0.34 & 0.42\nl
S Ori 39 & 05 38 32.40 & -02 29 58.0 & 15.47 & 14.85 & 14.43 & 13.74 & 0.62 & 0.42 & 0.69\nl
S Ori 40 & 05 37 36.40 & -02 41 57.0 & 15.49 & 14.94 & 14.59 & 13.86 & 0.55 & 0.35 & 0.73\nl
Cha H$\alpha$ 1 & 11 07 17.00 & -77 35 54.0 & 13.55 & 12.78 & 12.28 & 11.39 & 0.77 & 0.50 & 0.89\nl
Cha H$\alpha$ 2 & 11 07 43.00 & -77 33 59.0 & 12.59 & 11.43 & 11.15 & 9.68 & 1.16 & 0.28 & 1.47\nl
Cha H$\alpha$ 3 & 11 07 52.90 & -77 36 56.0 & 12.46 & 11.64 & 11.11 & 10.50 & 0.82 & 0.53 & 0.61\nl
Cha H$\alpha$ 4 & 11 08 19.60 & -77 39 17.0 & 12.20 & 11.42 & 11.04 & 10.58 & 0.78 & 0.38 & 0.46\nl
Cha H$\alpha$ 5 & 11 08 25.60 & -77 41 46.0 & 12.14 & 11.21 & 10.76 & 10.16 & 0.93 & 0.45 & 0.60\nl
Cha H$\alpha$ 6 & 11 08 40.20 & -77 34 17.0 & 12.43 & 11.61 & 11.09 & 10.23 & 0.82 & 0.52 & 0.86\nl
Cha H$\alpha$ 7 & 11 07 38.40 & -77 35 30.0 & 13.89 & 13.00 & 12.51 & 11.29 & 0.89 & 0.49 & 1.22\nl
Cha H$\alpha$ 8 & 11 07 47.80 & -77 40 08.0 & 12.92 & 12.12 & 11.63 & 10.93 & 0.80 & 0.49 & 0.70\nl
Cha H$\alpha$ 9 & 11 07 19.20 & -77 32 52.0 & 13.92 & 12.59 & 11.82 & 10.70 & 1.33 & 0.77 & 1.12\nl
Cha H$\alpha$ 10 & 11 08 25.60 & -77 39 30.0 & 14.41 & 13.68 & 13.27 & 12.51 & 0.73 & 0.41 & 0.76\nl
Cha H$\alpha$ 11 & 11 08 30.80 & -77 39 19.0 & 14.72 & 13.98 & 13.57 & 12.79 & 0.74 & 0.41 & 0.78\nl
Cha H$\alpha$ 12 & 11 06 37.50 & -77 43 07.0 & 13.20 & 12.42 & 12.00 & 11.17 & 0.78 & 0.42 & 0.83\nl
CHXR 73 & 11 06 28.90 & -77 37 33.0 & 13.00 & 11.32 & 10.79 & 10.11 & 1.68 & 0.53 & 0.68\nl
CHXR 74 & 11 06 57.40 & -77 42 10.4 & 11.55 & 10.57 & 10.23 & 9.75 & 0.98 & 0.34 & 0.48\nl
CHXR 78C & 11 08 54.60 & -77 32 12.0 & 12.41 & 11.64 & 11.28 & 10.66 & 0.77 & 0.36 & 0.62\nl
Gizis 1 & 11 39 51.10 & -31 59 21.0 & 12.62 & 12.16 & 11.57 & 10.97 & 0.46 & 0.59 & 0.60\nl
Gizis 2 & 12 07 33.40 & -39 32 54.0 & 13.02 & 12.57 & 12.02 & 11.38 & 0.45 & 0.55 & 0.64\nl
%UScoCTIO 85 & 15 54 03.51 & -23 12 31.0 & 12.97 & 12.34 & 11.91 & 11.18 & 0.63 & 0.43 & 0.73\nl
UScoCTIO 100 & 16 02 04.13 & -20 50 41.5 & 12.83 & 12.20 & 11.78 & 11.52 & 0.63 & 0.42 & 0.26\nl
UScoCTIO 104 & 15 57 12.66 & -23 43 45.3 & 13.52 & 12.88 & 12.52 & 11.87 & 0.64 & 0.36 & 0.65\nl
UScoCTIO 109 & 16 01 19.10 & -23 06 38.6 & 13.64 & 13.03 & 12.66 & 12.11 & 0.61 & 0.37 & 0.55\nl
UScoCTIO 112 & 16 00 26.57 & -20 56 32.0 & & & 12.81 & 12.06 & & & 0.75\nl
UScoCTIO 128 & 15 59 11.20 & -23 37 59.0 & 14.41 & 13.72 & 13.27 & 12.20 & 0.69 & 0.45 & 1.07\nl
UScoCTIO 130 & 15 59 43.56 & -20 14 38.1 & 14.29 & 13.54 & 13.12 & 12.49 & 0.75 & 0.42 & 0.63\nl
UScoCTIO 132 & 15 59 37.74 & -22 54 09.5 & 14.31 & 13.55 & 13.08 & 12.21 & 0.76 & 0.47 & 0.87\nl
UScoCTIO 137 & 15 56 47.87 & -23 47 44.0 & 15.60 & 14.92 & 14.41 & 14.04 & 0.68 & 0.51 & 0.37\nl
CRBR 14 & 16 26 19.07 & -24 26 12.0 & 15.20 & 13.50 & 12.34 & 10.87 & 1.70 & 1.16 & 1.47\nl
CRBR 15 & 16 26 19.05 & -24 24 16.8 & 16.30 & 13.90 & 11.94 & 9.62 & 2.40 & 1.96 & 2.32\nl
GY 5 & 16 26 21.70 & -24 26 02.0 & 12.70 & 11.60 & 10.94 & 10.07 & 1.10 & 0.66 & 0.87\nl
GY 10 & 16 26 22.30 & -24 23 54.0 & 15.80 & 13.50 & 12.24 & 11.08 & 2.30 & 1.26 & 1.16\nl
GY 11 & 16 26 22.40 & -24 24 08.0 & 16.50 & 15.40 & 14.15 & & 1.10 & 1.25 & \nl
GY 59 & 16 26 31.30 & -24 25 32.0 & 14.70 & 12.80 & 11.68 & 10.73 & 1.90 & 1.12 & 0.95\nl
GY 64 & 16 26 32.60 & -24 26 38.0 & 16.50 & 14.70 & 13.34 & $>$ 12.80 & 1.80 & 1.36 & $<$ 0.54\nl
GY 141 & 16 26 51.40 & -24 32 44.0 & 15.10 & 14.40 & 13.90 & & 0.70 & 0.50 & \nl
GY 202 & 16 27 06.00 & -24 28 36.0 & 16.76 & 14.52 & 13.01 & 11.71 & 2.24 & 1.51 & 1.30\nl
\enddata
\tablenotetext{a}{$J,H$ magnitudes for IC 348 sources from Luhman (1999), $J,H,K$ magnitudes for Taurus
sources from Brice\~{n}o et al. (2002), $J,H,K$ magnitudes for Chamaeleon sources 1 - 7, and 9 - 11 taken
from Comer\'{o}n, Neuh\"{a}user, \& Kaas (2000), $J,H$ magnitudes for Upper Sco 128, and 130 taken from
Ardila, Mart\'{i}n, \& Basri (2000).}
\end{deluxetable}

\clearpage
\begin{deluxetable}{llrr}
\footnotesize
\tablecaption{Spectral Type, Infrared Excess, and W$_{\lambda}$(H$\alpha$)
\label{table10}}
\tablewidth{0pt}
\tablehead{Source & ST\tablenotemark{a} & E(K-L$^\prime$)$_{0}$ & W$_{\lambda}$(H$\alpha$)(\AA)\tablenotemark{a}}
\startdata
IC348 165 & M5.25 & 0.48 & 66.0\nl
IC348 256 & M5.75 & 0.40 & 23.0\nl
IC348 286 & M5.75 & 0.08 & 6.9\nl
IC348 363 & M8 & -0.15 & 3.8\nl
IC348 367 & M4.75 & 0.28 & 6.8\nl
IC348 478 & M6.25 & -0.21 & 100.0\nl
KPNO-Tau 1 & M8.5 & 0.26 & 15.0\nl
KPNO-Tau 2 & M7.5 & 0.02 & 12.0\nl
KPNO-Tau 3 & M6 & 0.44 & 130.0\nl
KPNO-Tau 4 & M9.5 & -0.04 & 150.0\nl
KPNO-Tau 5 & M7.5 & 0.07 & 30.0\nl
KPNO-Tau 6 & M8.5 & 0.41 & 350.0\nl
KPNO-Tau 7 & M8.25 & 0.22 & 300.0\nl
KPNO-Tau 8 & M5.75 & 0.12 & 17.5\nl
KPNO-Tau 9 & M8.5 & 0.49\nl
S Ori 12 & M6 & 0.37 & 6.5\nl
S Ori 17 & M6 & -0.20 & 5.5\nl
S Ori 25 & M6.5 & 0.13 & 45.0\nl
S Ori 29 & M6 & -0.06 & 28.0\nl
S Ori 39 & M6.5 & 0.19 & 5.1\nl
S Ori 40 & M7 & 0.23 & 30.0\nl
Cha H$\alpha$ 1 & M7.5 & 0.23 & 99.0\nl
Cha H$\alpha$ 2 & M6.5 & 0.82 & 32.0\nl
Cha H$\alpha$ 3 & M7 & -0.08 & 13.0\nl
Cha H$\alpha$ 4 & M6 & -0.05 & 11.0\nl
Cha H$\alpha$ 5 & M6 & 0.01 & 11.0\nl
Cha H$\alpha$ 6 & M7 & 0.33 & 76.0\nl
Cha H$\alpha$ 7 & M8 & 0.54 & 45.0\nl
Cha H$\alpha$ 8 & M6.5 & 0.16 & 9.0\nl
Cha H$\alpha$ 9 & M6 & 0.54 & 16.0\nl
Cha H$\alpha$ 10 & M7.5 & 0.15 & 9.0\nl
Cha H$\alpha$ 11 & M8 & 0.12 & 23.0\nl
Cha H$\alpha$ 12 & M7: & 0.27 & 20.0\nl
CHXR 73 & M4.5: & -0.02 & \nl
CHXR 74 & M4.5 & -0.03 & 13.0\nl
CHXR 78C & M5.5 & 0.08 & 3.2\nl
Gizis 1 & M8 & -0.06 & 10.0\nl
Gizis 2 & M8 & -0.02 & 300.0\nl
%UScoCTIO 85 & M6 & 0.25 & 6.0\nl
UScoCTIO 100 & M7 & -0.24 & 13.1\nl
UScoCTIO 104 & M5 & 0.23 & 9.4\nl
UScoCTIO 109 & M6 & 0.07 & 19.0\nl
UScoCTIO 112 & M5.5 & 0.30 & 9.5\nl
UScoCTIO 128 & M7 & 0.57 & 24.8\nl
UScoCTIO 130 & M7.5 & 0.03 & 8.4\nl
UScoCTIO 132 & M7 & 0.37 & $<$1.0\nl
UScoCTIO 137 & M7 & -0.13 & \nl
CRBR 14 & M5.5 & 0.55 & \nl
CRBR 15 & M5 & 1.15 & \nl
GY 5 & M5.5 & 0.21 & 64.9\nl
GY 10 & M6.5 & 0.06 & \nl
GY 11 & M6.5 & & \nl
GY 59 & M5 & 0.02 & \nl
GY 64 & M7 & $<$0.04 & \nl
GY 141 & M8.5 & & 13.4\nl
GY 202 & M6.5 & 0.21 & \nl
\enddata
\tablenotetext{a}{Spectral types and W$_{\lambda}$(H$\alpha$) for IC 348 source 478 taken from Luhman (1999),
W$_{\lambda}$(H$\alpha$) for remaining IC 348 sources from Jayawardhana, Mohanty, \& Basri (2003); spectral
types and W$_{\lambda}$(H$\alpha$) for Taurus, Sigma Orionis, Chamaeleon, and Upper Sco sources from
Brice\'{n}o et al. (2002), B\'{e}jar, Zapatero Osorio, \& Rebolo (1999), Comer\'{o}n, Neuh\"{a}user, \& Kaas
(2000), and Ardila, Mart\'{i}n, \& Basri (2000) respectively. Spectral types and W$_{\lambda}$(H$\alpha$) for
Gizis 1 and 2 from Gizis (2002). Spectral types and W$_{\lambda}$(H$\alpha$) for $\rho$ Oph sources from
Luhman \& Rieke (1999) and Jayawardhana, Mohanty, \& Basri (2002) respectively.}
\end{deluxetable}

\clearpage
\begin{deluxetable}{lll}
\footnotesize
\tablecaption{Disk Fractions for Surveyed Regions
\label{table11}}
\tablewidth{0pt}
\tablehead{Region & Fraction & Estimated Age}
\startdata
$\rho$ Oph & 6/9 (67\% $\pm$ 27\%) & $\lesssim$1 Myr\nl 
IC 348 & 3/6 (50\% $\pm$ 29\%)     & $\sim$1 Myr\nl
Taurus & 5/9 (56\% $\pm$ 25\%)     & $\sim$1-3 Myr\nl
Cha I & 6/15 (40\% $\pm$ 16\%) & $\sim$1-3 Myr\nl
Upper Sco & 4/8 (50\% $\pm$ 25\%) & $\sim$3-5 Myr\nl
$\sigma$ Ori & 2/6 (33\% $\pm$ 24\%)  & $\sim$5-7 Myr\nl
TW Hya & 0/2                       & $\sim$10 Myr\nl
\enddata
\end{deluxetable}

\end{document}